\documentclass[10pt, draftclsnofoot, twocolumn]{IEEEtran}
\IEEEoverridecommandlockouts 
\usepackage{epsfig,latexsym}
\usepackage{float}
\usepackage{indentfirst}
\usepackage{amsmath}
\usepackage{amssymb}
\usepackage{times}
\usepackage{subfigure}
\usepackage{pifont}
\usepackage{psfrag}
\usepackage{cite}
\usepackage{url}
\usepackage{lastpage}
\usepackage{caption}
\usepackage{bm}
\usepackage{color}
\usepackage{fancyhdr}
\usepackage{balance}
\usepackage{fancyhdr,lastpage}
\usepackage{hyperref}
\usepackage{epstopdf}
\usepackage{algorithm}
\usepackage{algorithmic}
\usepackage{booktabs}
\usepackage{multirow}
\usepackage{colortbl}
\usepackage{color}
\definecolor{tabcolor}{rgb}{.105,.110,.113}
\usepackage{array}
\usepackage[T1]{fontenc}
\begin{document}
\title{Intelligent Reflective Surface Assisted Integrated Sensing and Wireless Power Transfer}
\vspace{-0.2in}
\author{\IEEEauthorblockN{  Zheng Li,~\emph{Graduate Student Member, IEEE}, Zhengyu Zhu,~\emph{Senior Member, IEEE}, Zheng Chu, Yingying Guan, De Mi,~\emph{Senior Member, IEEE}, Fan Liu,~\emph{Member, IEEE}, and Lie-Liang Yang,~\emph{Fellow, IEEE}		}
\thanks{(Corresponding author: Zhengyu Zhu)}
\thanks{Z. Li and Z. Zhu are with the School of Electrical and Information Engineering, Zhengzhou University, Zhengzhou, 450001, China, and Z. Zhu is also with National Mobile Communications Research Laboratory, Southeast University, Nanjing 210019, China. (e-mail: stones\_li@outlook.com, iezyzhu@zzu.edu.cn).
Z. Chu is with the 5GIC \& 6GIC, Institute for Communication Systems (ICS), University of Surrey, Guildford GU2 7XH, UK. (e-mail: andrew.chuzheng7@gmail.com).
Y. Guan is with the School of Computer Science and Engineering, Northeastern University, Shenyang 110819, China (e-mail: yingyingguan@foxmail.com)
D. Mi is with the School of Computing and Digital Technology, Birmingham City University, Birmingham, B4 7XG, U.K. (email: de.mi@bcu.ac.uk).
Fan Liu is with the Department of Electronic and Electrical Engineering, Southern University of Science and Technology, Shenzhen 518055, China (e-mail: liuf6@sustech.edu.cn).
L.-L. Yang is with the School of Electronics and Computer Science, University of Southampton, Southampton SO17 1BJ, U.K. (e-mail:lly@ecs.soton.ac.uk).
	}
}
\maketitle
\thispagestyle{empty}
\begin{abstract}
 Wireless sensing and wireless energy are enablers to pave the way for smart transportation and a greener future. In this paper, an intelligent reflecting surface (IRS) assisted integrated sensing and wireless power transfer (ISWPT) system is investigated, where the transmitter in transportation infrastructure networks sends signals to sense multiple targets and simultaneously to multiple energy harvesting devices (EHDs) to power them.
 In light of the performance tradeoff between energy harvesting and sensing, we propose to jointly optimize the system performance via optimizing the beamforming and IRS phase shift.
However, the coupling of optimization variables makes the formulated problem non-convex. Thus, an alternative optimization approach is introduced and based on which two algorithms are proposed to solve the problem.  Specifically, the first one involves a semi-definite program technique, while the second one features a low-complexity optimization algorithm based on successive convex approximation and majorization minimization. Our simulation results validate the proposed algorithms and demonstrate the advantages of using IRS to assist wireless power transfer in ISWPT systems.
\end{abstract}
\begin{IEEEkeywords}
 Intelligent reflecting surface, integrated sensing and wireless power transfer, semi-definite program, successive convex approximation,  majorization minimization.
\end{IEEEkeywords}
\IEEEpeerreviewmaketitle
\setlength{\baselineskip}{1\baselineskip}
\newtheorem{definition}{Definition}
\newtheorem{fact}{Fact}
\newtheorem{assumption}{Assumption}
\newtheorem{theorem}{Theorem}
\newtheorem{lemma}{Lemma}
\newtheorem{corollary}{Corollary}
\newtheorem{proposition}{Proposition}
\newtheorem{example}{Example}
\newtheorem{remark}{Remark}
\section{Introduction}
The Beyond 5G networks are expected to feature various capabilities and integrate various network functions, to satisfy the requirement for challenging sectors, e.g., smart transportation.
On one hand, the integration of sensing and communications within the transportation infrastructures is a technological momentum towards the evolution of the transportation sector\cite{8999605}.
In the past, sensing is usually studied in an independent manner to be used in specific areas.
However, due to the stimulating demand for the smart transportation with vertical user scenarios, the number of communication scenarios requiring high-precision sensing support is increasing.
On the other hand, using massive intelligent IoT devices helps the smart transportation become greener, more digital and sustainable, while power supply in large-scale IoT networks has always been a concern.
For example, traditional devices in smart IoT systems, are usually powered by batteries, which have limited  energy and also cause environmental pollution.
By contrast, the devices with energy harvesting capability, such as those that can harvest energy from  radio frequency (RF) signals in the environment, have a sustainable operation ability and are environmentally-friendly, based on which green and efficient IoT systems are possible\cite{8187650}.
Furthermore, to achieve high-efficiency wireless power transfer (WPT), accurate positioning of the target devices is desired.
Therefore, the transmitters in the smart transportation infrastructure networks with both sensing and WPT capabilities have great application potential in bolstering future transportation systems\cite{10077119}.

However, the practical complex propagation conditions can be the showstopper of the integration of sensing with WPT\cite{9566705}. Specifically, the path-loss of wireless channels can severely affect the efficiency of energy harvesting and thus limit the application of WPT. For sensing, a blockage of direct links can fail the target acquisition.
To mitigate these problems, an intelligent reflecting surface (IRS) has been introduced\cite{9838546}. An IRS, which is equipped with a large number of controllable reflective elements, is able to enhance the wireless channel by modulating the phases of the incident signals, thus providing an additional link to improve the performance of wireless communication\cite{10124870}.
Therefore, incorporating IRS into integrated sensing and wireless power transfer (ISWPT) systems provides a promising solution to combat the path-loss and blockage problems in wireless systems.

In recent works, the feasibility of integrating sensing and communication was verified in \cite{8288677} by fusing sensing into a wireless system, and the integration of sensing and WPT has been considered in \cite{9928277}.
The authors in \cite{9349191} demonstrated the benefits of the IRS-assisted WPT and the advantages of using the IRS for performance enhancement in WPT systems.
In \cite{9771801}, the authors investigated an IRS-assisted sensing communication by designing the base station (BS) beamforming and IRS phase shift with the objective to achieve target sensing in the case of the line of sight (LoS) blockage. In \cite{9769997},  the potential of applying IRS to a dual-function radar communication system to improve radar sensing and communication capabilities was studied, where a multi-antenna BS performs both radar sensing and multi-user communication on the same hardware platform.
The authors in \cite{chen2022isac} considered sensing to simultaneous wireless information and power transfer system in the same platform, i.e. sharing the same antenna and transmit power resources.
We consider a practically important case where a multiple-antenna transmitter generates a waveform to perform both sensing and WPT functions simultaneously with the assistance of an IRS.
 We propose to deal with this via a trade-off controlled problem solved by jointly optimizing the transmit beamforming vector and the IRS phase shifts. Our studies to the main contributions of this paper as summarized below.

\begin{itemize}
  \item We investigate an IRS-assisted ISWPT system that simultaneously provides sensing and WPT for multiple targets and energy harvesting devices (EHDs). To optimize the overall system performance, we formulate an optimization problem to optimize the beamforming and IRS phase shift, where a trade-off factor is introduced to obtain the desired performance trade-off between sensing and WPT.

  \item To solve the formulated non-convex problem with coupling optimization variables, a semi-definite program (SDP) technique is designed under the framework of alternating optimization (AO). Then, to obtain the approximate solvable sub-problems, the low-complexity algorithms based on successive convex approximation (SCA) and majorization minimization (MM) algorithm are proposed. Finally, we verify the effectiveness of the proposed algorithms and demonstrate  the advantages of the IRS by simulation.
\end{itemize}

\textbf{Notations}: Lowercase boldface and uppercase boldface symbols indicate vectors and matrices, respectively. ${\mathbf{A}}\underline  \succ  \bm{0}$ means that ${\mathbf{A}}$ is a semi-positive definite matrix. ${{\bf{A}}^H}$, ${{\bf{A}}^T}$, and ${{\bf{A}}^*}$ represent the conjugate transpose, transpose, and the conjugate, respectively.
${{\lambda _{\max }}({\bf{A}})}$ is the maximum eigenvalue of $\mathbf{A}$.
${diag}\left( {\mathbf{a}} \right)$ denotes a diagonal matrix whose diagonal elements are vector ${\bf{a}}$. ${conj(}{\mathbf{a}}{\text{)}}$ and $\arg \left( {\mathbf{a}} \right)$ stand for the conjugate and the phase of a complex vector  ${\mathbf{a}}$, respectively. $ \odot $ and $ \otimes $ denote the Hadamard and Kronecker products. $\Re \left(  \cdot  \right)$ is the real part. ${\mathbf{I}}$ is the identify matrix.

\section{System Model}
\pagestyle{headings}
\setcounter{page}{1}
\pagenumbering{arabic}
\begin{figure}[h]
  \centering
  \includegraphics[width= 8 cm,height=6cm]{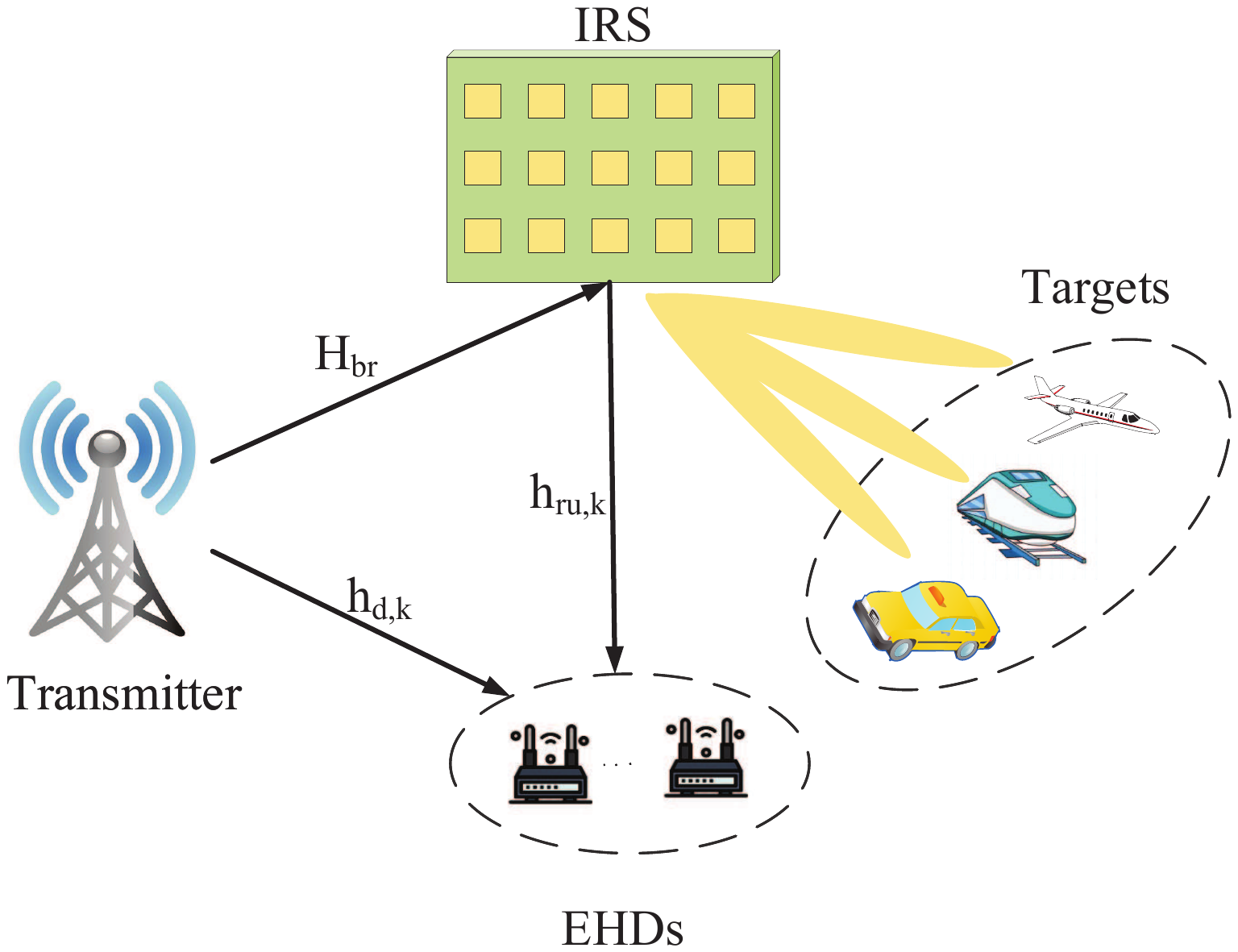}\\
  \caption{ system model}
  \vspace{-0.2in}
\end{figure}

Fig. 1 illustrates the considered IRS-assisted ISWPT system, which consists of a fixed transmitter equipped with $N$ antennas, an IRS with $L$ reflecting elements, $K$ EHDs with single antenna, and $M$ interested moving targets in a smart transportation setup. The transmitter sends radar signals to realize the sensing of the targets, and the EHDs can also harvest the energy from the radar signals. Thus, in the considered ISWPT system, no distinction between radar signals and WPT signals is needed. Thus, the transmitted signal can be expressed as
\begin{equation}
 {\bf{x}} =  {{{\bf{w}}}{s}},
\end{equation}
where ${\mathbf{w}} \in {\mathbb{C}^{N \times 1}}$ is the  transmitter beamforming vector, and ${s}$ is the independently and uncorrelated radar symbol, which satisfies $\mathbb{E}\left[ {{{\left| s \right|}^2}} \right] = 1$.
The covariance matrix of the transmitted signal is $ vbn{\mathbf{W}} \triangleq \mathbb{E}\left[ {{\mathbf{x}}{{\mathbf{x}}^H}} \right] = {\mathbf{w}}{{\mathbf{w}}^H}$.
Considering the per-antenna transmit power constraints, ${\mathbf{W}}$ should satisfy
\begin{equation}
  {{\mathbf{W}}(n,n)} = {P_0}/N,
\end{equation}
where ${P_0}$ is the power budget of the transmitter.

We assume that the LoS,  between the transmitter and targets are blocked. Thus, an IRS is needed to assist the transmitter to generate a virtual LoS link to achieve the sensing.
We assume that the IRS has a uniform linear array of $L$ elements, and the phase shift of the IRS is denoted as $\bm\Theta  =\operatorname{diag}\left(e^{j \alpha_{1}}, \ldots, e^{j \alpha_{L}}\right) \in {\mathbb{C}^{L \times L}}, \alpha_{l}  \in[-\pi, \pi], l = 1,...,L$. Accordingly, the beampattern of the IRS towards the interested targets is adopted as the sensing performance metric. The steering vector at the IRS for the angle of deviation $\theta_m$ where the $m$-th target located is

\begin{equation}
  {\bf{a}}({\theta _m}) = \left[ {1,{{\mathop{\rm e}\nolimits} ^{j2\pi \Delta \sin ({\theta _m})}},...,{{\mathop{\rm e}\nolimits} ^{j2\pi (L - 1)\Delta \sin ({\theta _m})}}} \right],
\end{equation}
where $\Delta$ is the spacing between two adjacent reflecting elements of the IRS. Thus, the beampattern from the IRS towards desired angle ${\theta _m}$ formulated as
\begin{equation}
\begin{gathered}
  \mathcal{P}({\theta _m}){\text{ }} = \mathbb{E}\left( {{{\left| {{{\mathbf{a}}}({\theta _m}){\mathbf{\Theta }}{{\mathbf{H}}_{br}}{\mathbf{x}}} \right|}^2}} \right)
  =  \mathbb{E}\left({\left| {{{{\mathbf{\hat h}}}_m}{\mathbf{x}}} \right|^2} \hfill\right)
  \\= \left( {{{\mathbf{a}}}({\theta _m}){\mathbf{\Theta }}{{\mathbf{H}}_{br}}} \right){\mathbf{W}}\left( {{\mathbf{H}}_{br}^H{\mathbf{\Theta }}_{}^H{\mathbf{a}^H}({\theta _m})} \right).{\text{ }} \hfill \\
\end{gathered}
\end{equation}
where ${{\mathbf{H}}_{br}} \in {\mathbb{C}^{L \times N}}$ is the channel vector from the transmitter to IRS, and ${\mathbf{\hat h}_m} = {{\mathbf{a}}}(\theta_m ){{\mathbf{\Theta }}}{{\mathbf{H}}_{br}}$ for similarity.

Concerning the EHDs, by combining the direct link and the cascaded link via IRS, the received signal of the $k$-th EHD can be given as
\begin{equation}
  {y_k} = \sqrt {{P_0}} \left( {{{\bf{h}}_{ru,k}}{\bf{\Theta }}{{\bf{H}}_{br}} + {{\bf{h}}_{d,k}}} \right){\bf{x}}  + {n_k},
\end{equation}
where ${{\mathbf{h}}_{ru,k}} \in {\mathbb{C}^{1 \times L}}$ and ${{\mathbf{h}}_{d,k}} \in {\mathbb{C}^{1 \times N}}$ are the channels from IRS to the $k$-th EHD and that from transmitter to the $k$-th EHD, respectively. $n_k$ is the additive white Gaussian noises. Hence, the harvested power of the $k$-th EHD can be expressed by\cite{9559408}
\begin{equation}
  {E_k} = \eta {P_0}{\left| {{{{\mathbf{\tilde h}}}_k}{\mathbf{w}}} \right|^2},
\end{equation}
where ${\mathbf{\tilde h}_k} = {{\mathbf{h}}_{ru,k}}{{\mathbf{\Theta }}}{{\mathbf{H}}_{br}} + {{\mathbf{h}}_{d,k}}$ and $\eta$ represents the efficiency of the energy harvesting.
\subsection{Problem Formulation}
Our optimization objective is to improve the overall performance of the ISWPT system with the assistance of IRS, for which the beampattern gain and harvested energy are jointed to form an optimization metric. To achieve the maximum performance gain, we design the beamforming vector and IRS phase shift. Specifically, to strike a good sensing performance and the WPT performance, a trade-off factor $ \rho$ is introduced to schedule the allocation of the two objective functions, by formulating an optimization as
\begin{subequations}\label{problem}
\begin{align}
 & \mathop {\max }\limits_{{{\mathbf{w}}},{\mathbf{\Theta }}} ~\rho {P_0}\sum\limits_{k = 1}^K {{E_k}}  + (1 - \rho )\sum\limits_{m = 1}^M {\mathcal{P}({\theta _m}){\text{ }}}  \hfill \\
 & ~{\text{ s.t.}} ~\left| {{\mathbf{w}}(n)} \right| = \sqrt {{P_0}/N} ,n = 1,...,N, \hfill \label{powerc}\\
 &~~~~~~ {\text{        }}{{\mathbf{\Theta }}} = \operatorname{diag} \left( {{e^{j\alpha _1}}, \ldots ,{e^{j\alpha _{L}}}} \right), \label{unit}\hfill
\end{align}
\end{subequations}
where \eqref{powerc} denotes the transmit power constraint per-antenna, and \eqref{unit} is the phase shift constraints.



\section{Proposed Solutions}
The mutual coupling of the optimization variables of problem beamforming $\mathbf{w}$ and IRS phase shift $\mathbf{\Theta }$ in \eqref{problem} makes the optimization non-convex and difficult to solve directly. To address this problem, we propose to use the AO algorithm and two suboptimal solutions to approximately solve the problem. Specifically, first suboptimal scheme is proposed based on the SDP technique to relax the problem. Then, a low-complexity algorithm is designed to optimize the beamforming $\mathbf{w}$ using  the SCA approximation, and the IRS phase shifts based on the MM algorithm.

\subsection{SDP Algorithm}\label{SDR1}
\subsubsection{Optimization of Beamforming Vector $\mathbf{w}$}
Given the IRS phase shifts $\mathbf{\Theta }$,
the problem \eqref{problem} can be equivalently rewritten as
\begin{subequations}\label{Beam1}
  \begin{align}
  &\mathop {\max }\limits_{{{\mathbf{W}}}}~ \rho \eta {P_0} \sum\limits_{k = 1}^K { {{\text{Tr}}({{{\mathbf{\tilde h}}}_k^H}{\mathbf{\tilde h}}_k{{\mathbf{W}}}) + (1-\rho)\sum\limits_{m = 1}^M { {{\text{Tr(}}{{{\mathbf{\hat h}}}_m^H}{\mathbf{\hat h}}_m{{\mathbf{W}}}{\text{) }}} }}}  \hfill \\
  &~{\text{ s.t.}}~{{\mathbf{W}}_{n,n}} = {P_0}/N, {\mathbf{W}}\underset{\raise0.3em\hbox{$\smash{\scriptscriptstyle-}$}}{ \succ } {\mathbf{0}},
  ~\text{rank}({{\mathbf{W}}}) = 1. \label{rankone}\hfill
\end{align}
\end{subequations}

By relaxing the rank-one constraint in \eqref{rankone}, the optimization problem \eqref{Beam1} is a standard convex problem, which can be solved using the existing optimization tools \cite{Boyd2004Convex}.

\subsubsection{Optimization of IRS Phase Shifts $\mathbf{\Theta }$}
After obtaining the beamforming vector $\mathbf{w}$, IRS phase shifts $\mathbf{\Theta }$ are then optimized. First, the following mathematical equivalent transformation is performed
\begin{subequations}
  \begin{align}
  & {E_k} = \eta {P_0}{\left| {({{\mathbf{h}}_{d,k}} + {\mathbf{v}}{\text{diag(}}{{\mathbf{h}}_{ru,k}}{\text{)}}{{\mathbf{H}}_{br}}){\mathbf{w}}} \right|^2} = \eta P{\left| {{\mathbf{v}}{{\mathbf{c}}_k} + {a_k}} \right|^2}, \hfill \\
  &\mathcal{P}({\theta _m}) = {\left| {{\mathbf{v}}{\text{diag(}}{{\mathbf{a}}^{\text{H}}}({{\mathbf{\theta }}_m})){{\mathbf{H}}_{br}}{\mathbf{w}}} \right|^{\text{2}}}{\text{ = }}{\left| {{\mathbf{v}}{{\mathbf{d}}_m}} \right|^{\text{2}}}, \hfill
\end{align}
\end{subequations}
where ${\mathbf{v}} = [{e^{j{\alpha _1}}}, \ldots ,{e^{j{\alpha _L}}}]$, ${{{\mathbf{c}}_{k}}} = {{\text{diag(}}{{\mathbf{h}}_{ru,k}}{\text{)}}{{\mathbf{H}}_{br}}{{\mathbf{w}}}}$, ${{a_{k}}} = {{{\mathbf{h}}_{d,k}}{{\mathbf{w}}}}$,
and ${{{\mathbf{d}}_{m}}}= {{\text{diag(}}{{\mathbf{a}}}({\theta _m})){{\mathbf{H}}_{br}}{{\mathbf{w}}}}$.

Correspondingly, the optimization problem \eqref{problem} can be rewritten as
\begin{subequations}\label{problem2}
\begin{align}
  &\mathop {\max }\limits_{\mathbf{v}} \rho \eta {P_0}\sum\limits_{k = 1}^K {{{\left| {{\mathbf{v}}{{\mathbf{c}}_k} + {a_k}} \right|}^2}}  + (1 - \rho )\sum\limits_{m = 1}^M {{{\left| {{\mathbf{v}}{{\mathbf{d}}_m}} \right|}^{\text{2}}}{\text{ }}}  \hfill \\
 & ~{\text{ s}}{\text{.t}}{\text{. }}\left| {{\mathbf{v}}(l)} \right| = 1,l = 1,...,L. \hfill
\end{align}
\end{subequations}

Upon defining ${\mathbf{F}} \triangleq  \left[ {\begin{array}{*{20}{c}}
  {{{\mathbf{F}}_{11}}}&{{{\mathbf{f}}_{12}}} \\
  {{{\mathbf{f}}_{21}}}&0
\end{array}} \right]$, where ${{\mathbf{F}}_{11}} = \rho\eta {P_0} \sum\limits_{k = 1}^K {{{\mathbf{c}}_k}{\mathbf{c}}_k^H}  + (1 - \rho )\sum\limits_{m = 1}^M {{{\mathbf{d}}_m}{\mathbf{d}}_m^H} $ and ${{\mathbf{f}}_{12}} = {\mathbf{f}}_{21}^* = \rho \eta {P_0}\sum\limits_{k = 1}^K {a_k^*{{\mathbf{c}}_k}} $,
 the problem \eqref{problem2} can now be described as
\begin{subequations}\label{problem3}
\begin{align}
 & \mathop {\max }\limits_{{\mathbf{\tilde v}}} ~{\mathbf{\tilde vF}}{{{\mathbf{\tilde v}}}^H} \hfill \\
 & {\text{ s}}{\text{.t}}{\text{. }}\left| {{\mathbf{\tilde v}}(l)} \right| = 1,l = 1,...,L + 1, \hfill
\end{align}
\end{subequations}
where ${\mathbf{\tilde v}} = [{\mathbf{v}},1]$.
Finally, by defining ${\mathbf{V}} \triangleq {\mathbf{\tilde v}^H}{{{\mathbf{\tilde v}}}}$, where ${\mathbf{V}}\underset{\raise0.3em\hbox{$\smash{\scriptscriptstyle-}$}}{ \succ } {\mathbf{0}}$ and ${\text{rank(}}{\mathbf{V}}){\text{ = 1}}$, we have ${\mathbf{\tilde vF}}{{{\mathbf{\tilde v}}}^H} = {\text{Tr(}}{\mathbf{FV}})$. Thus, with the aid of the SDP technique, the problem \eqref{problem3} can be rewritten as
\begin{subequations}\label{problem4}
  \begin{align}
  &\mathop {\max }\limits_{\mathbf{V}}~ {\text{Tr}}\left( {{\mathbf{FV}}} \right) \hfill \\
  &~~{\text{s}}{\text{.t}}{\text{.   }}~{\mathbf{V}}\underset{\raise0.3em\hbox{$\smash{\scriptscriptstyle-}$}}{ \succ } {\mathbf{0}}, \hfill \\
  &~~~~~~~{[{\mathbf{V}}]_{l,l}} = 1,l = 1,...,L + 1, \hfill \\
  &~~~~~~~{\text{rank(}}{\mathbf{V}}){\text{ = 1}}. \label{rank12}
\end{align}
\end{subequations}

Again, by relaxing the rank-one constraint in \eqref{rank12}, the problem \eqref{problem4} is a SDP problem, which can be solved by the optimization tools, such as CVX.

\subsection{Low Complexity Algorithm}\label{LC1}
To reduce the computational complexity, we propose a low-complexity algorithm, in which the beamforming $\mathbf{w}$ is designed by the SCA algorithm and the IRS phase shift $\mathbf{\Theta }$ is calculated by the MM algorithm.
\subsubsection{Optimization of Beamforming Vector $\mathbf{w}$}
Given the IRS phase shift $\mathbf{\Theta }$, the problem \eqref{problem} with respect to $\mathbf{w}$ can be reformulated as
\begin{equation}\label{problem5}
 \begin{gathered}
  \mathop {\max }\limits_{\mathbf{w}} {\sum\limits_{k = 1}^K {\left| {\sqrt {\rho \eta {P_0}} {{{\mathbf{\tilde h}}}_k}{\mathbf{w}}} \right|} ^2} + {\sum\limits_{m = 1}^M {\left| {\sqrt {1 - \rho } {{{\mathbf{\hat h}}}_m}{\mathbf{w}}} \right|} ^2} \hfill \\
  {\text{ s.t.}}~\left| {{\mathbf{w}}(n)} \right| = \sqrt {{P_0}/N} ,n = 1,...,N. \hfill \\
\end{gathered}
\end{equation}

The objective function in \eqref{problem5} is convex, which inspire us adopt the SCA algorithm for its solution. As shown in \cite{9349191}, for a given local point, any convex function is lower-bound globally by its first-order Talyor expansion at any feasible point. Thus,  the objective function of the original problem can be replaced by an approximation point near a feasible point in each step, and then, the resulting approximation problem can be solved to obtain an approximation point for the next iteration.

In detail, when a feasible point ${\bf{\hat w}}$, we have

\begin{equation}\label{144}
 \begin{split}
&\sum\limits_{k = 1}^K {{{\left| {\sqrt {\rho \eta } {{{\bf{\tilde h}}}_k}{\bf{w}}} \right|}^2}}  + \sum\limits_{m = 1}^M {{{\left| {\sqrt {(1 - \rho )} {{{\bf{\hat h}}}_m}{\bf{w}}} \right|}^2}} \\
& \ge 2\Re \left\{ {{{\bf{w}}^H}{\bf{H\hat w}}} \right\} - {\bf{\hat w}}_{}^H{\bf{H\hat w}},
\end{split}
\end{equation}
where ${\bf{H}} = {\bf{\tilde H}} + {\bf{\bar H}}$, with ${\bf{\tilde H}} = \sum\limits_{k = 1}^K {\rho \eta {P_0} } {{{\bf{\tilde h}}}_k^H}{\bf{\tilde h}}_k$, and ${\bf{\bar H}} = \sum\limits_{m = 1}^M {(1 - \rho )} {{{\bf{\hat h}}}_m^H}{\bf{\hat h}}_m$. The equality in \eqref{144} holds at point ${\mathbf{w}} = {\mathbf{\hat w}}$. Then, maximizing this approximation function under the constraint of \eqref{powerc}, it is easy to see that the updated beamforming vector is given by
\begin{equation}
 {{\mathbf{w}}^{opt}} = \sqrt {P/N} \exp \left( {j\arg ({\mathbf{{\bf H}\hat w}})} \right).
\end{equation}
\subsubsection{Optimization of  IRS Phase Shift $\mathbf{\Theta }$}
Given a feasible point $\mathbf{w}$ as obtained above, the problem \eqref{problem2} can be reformulated as
\begin{subequations}\label{problem6}
  \begin{align}
  &\mathop {\min }\limits_{\mathbf{v}} {\mathbf{vD}}{{\mathbf{v}}^H} - 2\Re \left( {{\mathbf{vc}}} \right) \\
  &{\text{ s}}{\text{.t}}{\text{. }}\left| {{\mathbf{v}}(l)} \right| = 1,l = 1,...,L, \hfill
\end{align}
\end{subequations}
where ${{\bf{ D}}} = - {{\mathbf{F}}_{11}}$ and ${\mathbf{c}} = {\mathbf{f}}_{12}^* $, the ${{\mathbf{F}}_{11}}$ and ${\mathbf{f}}_{12}$ are defined above (11).

The quadratic problem \eqref{problem6} by can be tackled the MM algorithm for deriving the optimal $\mathbf{v}$. The MM algorithm minimizes an objective function by decomposing it into a series of surrogate problems in which the objective functions are feasible convex functions in different iteration steps. Specifically, when given a feasible ${{\mathbf{v}}^{(i)}}$ in the $i$-th iteration, the objective function \eqref{problem6} satsifies
\begin{equation}\label{1777}
  \begin{split}
 &{\mathbf{vD}}{{\mathbf{v}}^H} - 2\Re \left( {{\mathbf{vc}}} \right) \\
 &\!\!\leqslant {\mathbf{vT}}{{\mathbf{v}}^H} - 2\Re \left( {{\mathbf{v}}\left[ {\left( {{\mathbf{T}} - {\mathbf{D}}} \right){{{\mathbf{\tilde v}}}^H} + {\mathbf{c}}} \right]} \right) + {\mathbf{\tilde v}}\left( {{\mathbf{T}} - {\mathbf{D}}} \right){{{\mathbf{\tilde v}}}^H} \hfill \\
 & \!\! = {\lambda _{\max }}\left( {\mathbf{D}} \right){\left\| {\mathbf{v}} \right\|^2}\! \!- \!\!2\Re \left( {{\mathbf{v}}\left[ {\left( {{\lambda _{\max }}\left( {\mathbf{D}} \right){{\mathbf{I}}_{L \times L}} \!\!- \!\!{\mathbf{D}}} \right){{{\mathbf{\tilde v}}}^H} \!+\! {\mathbf{c}}} \right]} \right) \!+\! \varsigma,  \hfill \\
\end{split}
\end{equation}
where ${\mathbf{T}} = {\lambda _{\max }}\left( {\mathbf{D}} \right){{\mathbf{I}}_{L \times L}}$, $\varsigma  = {\mathbf{\tilde v}}\left[ {\left( {{\lambda _{\max }}\left( {\mathbf{D}} \right){{\mathbf{I}}_{L \times L}} - {\mathbf{D}}} \right){{{\mathbf{\tilde v}}}^H} + {\mathbf{c}}} \right]$, and ${\mathbf{\tilde v}}$ obtained in the previous iteration represents the approximated solution to ${\mathbf{ v}}$. With the aid of \eqref{1777}, the problem \eqref{problem6} can be approximated as
\begin{subequations}\label{problem8}
  \begin{align}
  &\mathop {\min }\limits_{\mathbf{v}} {\text{ }}{\lambda _{\max }}\left( {\mathbf{D}} \right){{\mathbf{I}}_{L \times L}} - 2\Re \left( {{\mathbf{v\tilde c}}} \right) \hfill \label{18a}\\
  &{\text{  s}}{\text{.t}}{\text{. }}\left| {{\mathbf{v}}(l)} \right| = 1,l = 1,...,L, \label{18b}\hfill
\end{align}
\end{subequations}
where ${\mathbf{\tilde c}} = \left( {{\lambda _{\max }}\left( {\mathbf{D}} \right){{\mathbf{I}}_{L \times L}} - {\mathbf{D}}} \right){{{\mathbf{\tilde v}}}^H} + {\mathbf{c}}$. From \eqref{18a}, we can know that under the constraint of \eqref{18b},  $\Re \left( {{\mathbf{v\tilde c}}} \right)$ is maximized is and only if  ${\mathbf{v}}$ and ${{\mathbf{\tilde c}}}$ are equal. Thus, the optimal solution of problem \eqref{problem8} can be derived to be
\begin{equation}
 {{\mathbf{v}}^{opt}} = \left[ {{e^{j\arg \left[ {{\bm{\gamma }}(1)} \right]}},...,{e^{j\arg \left[ {{\bm{\gamma }}(L)} \right]}}} \right],
\end{equation}
where ${\bm{\gamma }} = \left( {{\lambda _{\max }}({\mathbf{D}}){{\mathbf{I}}_{L \times L}} - {\mathbf{D}}} \right){\mathbf{\tilde v}} + {\mathbf{c}}$.

\subsection{Complexity Analysis}
The computational complexities for solving \eqref{Beam1} and \eqref{problem4} are $\mathcal{O}\left( {{N^{6.5}}} \right)$ and $\mathcal{O}\left( {{L^{6.5}}} \right)$, respectively. Hence, the total complexity of the SDR algorithm is $\mathcal{O}\left( {{I_{{\text{sdr}}}}\left( {{N^{6.5}} + {L^{6.5}}} \right)} \right)$, where ${{I_{{\text{sdr}}}}}$ is the maximum number of  iterations. The computational complexity to update ${{\mathbf{w}}^{opt}}$ in \eqref{144} is $\mathcal{O}\left( {{N^2}} \right)$, and the computational
complexity of the MM algorithm is given as $\mathcal{O}\left( {{L^3} + {I_m}{L^2}} \right)$.
Therefore, the total complexity of our proposed low-complexity (LC) algorithm is  $\mathcal{O}\left( {{I_{lc}}(N^2 + {L^3} + {I_m}{L^2})} \right)$, where ${I_m}$ and ${I_{lc}}$ are the maximum number of iterations of the MM and LC algorithms.

\section{Results and Discussion}
In this section, we provide some numerical results to demonstrate the effectiveness of the proposed algorithms and show the benefits of IRS for enhancing the performance of both energy harvesting and beampattern gain. In our studies, the channel path loss follows ${P_l} = \varsigma {d^{ - \chi }}$, where $\varsigma = -10$ dB, $d$ is the distance, and $\chi $ is the path loss exponent. The distance between transmitter and IRS, that between IRS and EHDs, and that between transmitter and EHDs are 30 m, 30 m, and 50 m, respectively. The path-loss exponent for transmitter to IRS,  IRS-EHDs, and transmitter-EHDs are set to 2.5, 2.5, and 3, respectively. We assume the Rician fading channel coefficients between transmitter and IRS, IRS and $k$-th EHD, and the transmitter and $k$-th EHD are modelled as ${{\bf{H}}_{br}} = \sqrt {\frac{{{K_1}}}{{{K_1} + 1}}} {\bf{H}}_{br}^{{\rm{LOS}}} + \sqrt {\frac{1}{{{K_1} + 1}}} {\bf{H}}_{br}^{{\rm{NLOS}}}$, ${{\bf{h}}_{ru,k}} = \sqrt {\frac{{{K_1}}}{{{K_1} + 1}}} {\bf{h}}_{ru,k}^{{\rm{LOS}}} + \sqrt {\frac{1}{{{K_1} + 1}}} {\bf{h}}_{ru,k}^{{\rm{NLOS}}}$, and ${{\bf{h}}_{d,k}} = \sqrt {\frac{{{K_1}}}{{{K_1} + 1}}} {\bf{h}}_{d,k}^{{\rm{LOS}}} + \sqrt {\frac{1}{{{K_1} + 1}}} {\bf{h}}_{d,k}^{{\rm{NLOS}}}$, respectively. $K_1$ is the Rician factor,  set to 6 dB. The LoS and non-LoS components are independent and identically distributed with zero mean and unit variance.  We set $N = 12$, $K=5$, $P_0 = 30$ dBm, and $\eta = 0.8$. The interested targets are located in the angles $ - 45{\rm{^\circ }},~0{\rm{^\circ }},~45{\rm{^\circ }}$  from the IRS.

\begin{figure}[h]\label{fig1}
  \centering
  \includegraphics[scale = 0.55]{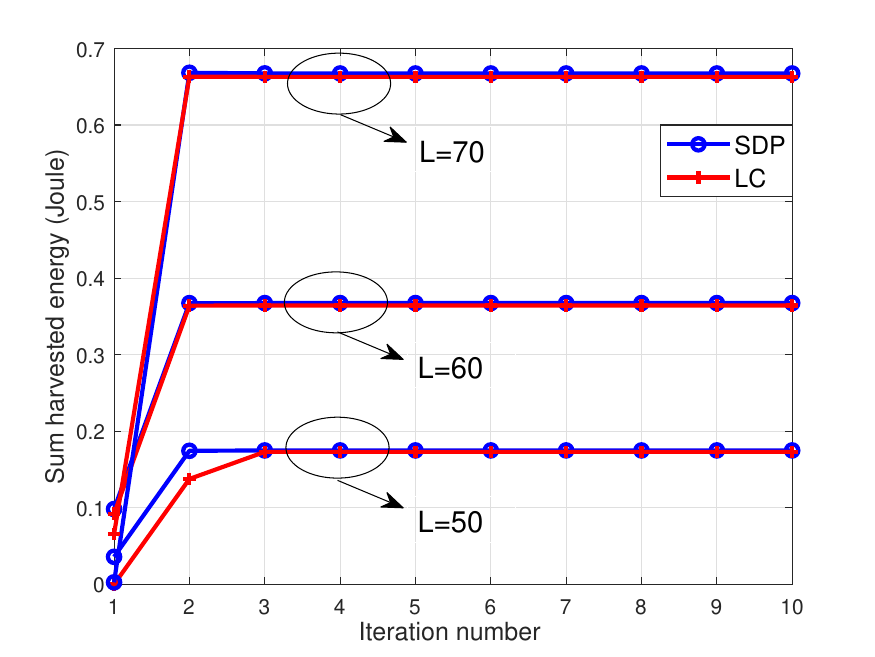}\\
  \caption{Convergence of the proposed SDP and LC algorithms}
\end{figure}

We first evaluated the convergence of the proposed SDP and LC algorithms. It can be seen from Fig. 2 that both algorithms converge very quickly when the IRS is equipped with different number of elements. Furthermore, the converged performance of both algorithms is closely consistent, typically 2-3 iterations.

\begin{figure}\label{fig1}
  \centering
  \includegraphics[scale = 0.55]{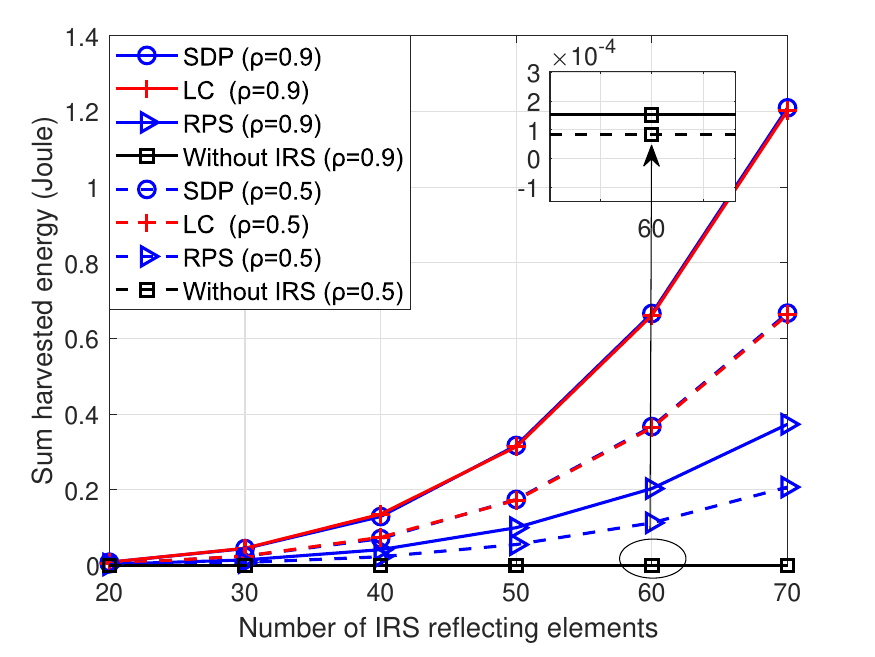}\\
  \caption{ Harvested energy with respect to $L$.}
\end{figure}

Then, we evaluate the effect of the trade off factor $\rho$ on the energy harvesting performance. From Fig. 3, it can be seen that when $\rho$ = 0.9, more resources are allocated for wireless energy harvesting, and the harvested energy is higher than that when $\rho$  = 0.5. As shown in Fig. 3, when $L$ = 20, the energy harvested by EHDs is very low.
The harvested energy increases significantly, when the number of the reflecting elements increases. Furthermore, the optimized IRS reflective phase scheme significantly outperforms the random phase scheme (RPS), which also validates the importance of the reflected phase optimization.

\begin{figure}\label{fig1}
  \centering
  \includegraphics[scale = 0.55]{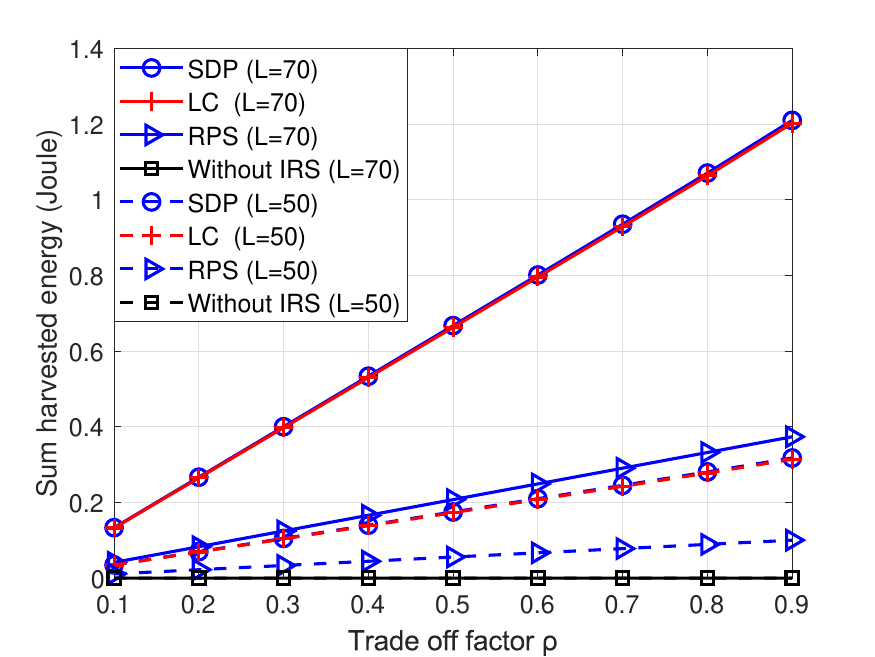}\\
  \caption{ Harvested energy with respect to $\rho$.}
\end{figure}

Next, we demonstrate the sum harvested energy versus the trade off factor $\rho$ is shown in Fig. 4. As expected, $\rho$ increases, more resources are allocated for energy harvesting, leading  to the improvement of the system performance in terms of SDP and LC energy harvesting. Again, it is shown that employing IRS is capable of significantly improving the energy harvesting, and our proposed SDP and LC algorithms are effective for performance improvement.
\begin{figure}
  \centering
  \includegraphics[scale = 0.55]{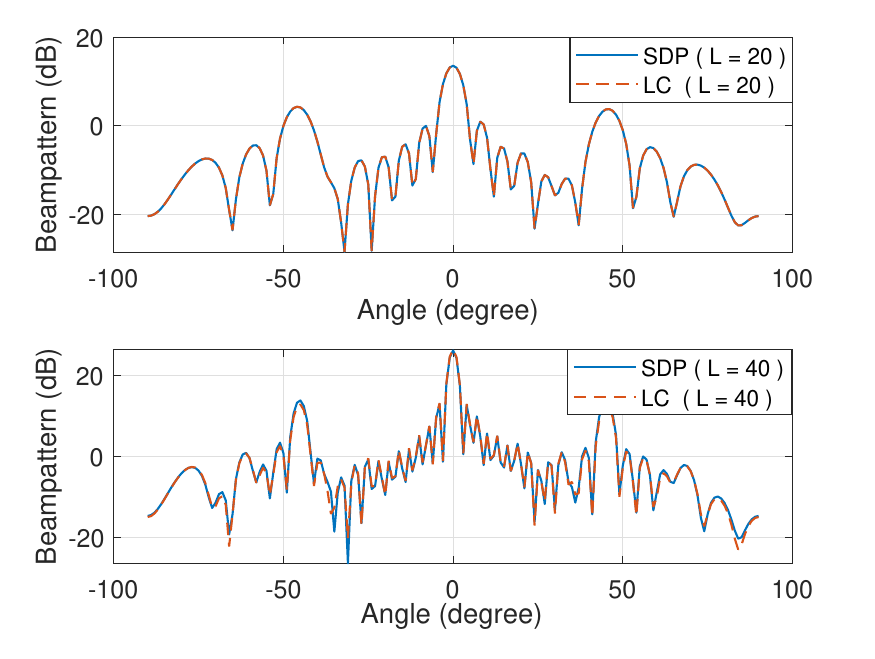}\\
  \caption{ Beampattern with respect to $L$.}
\end{figure}
\begin{figure}
  \centering
  \includegraphics[scale = 0.58]{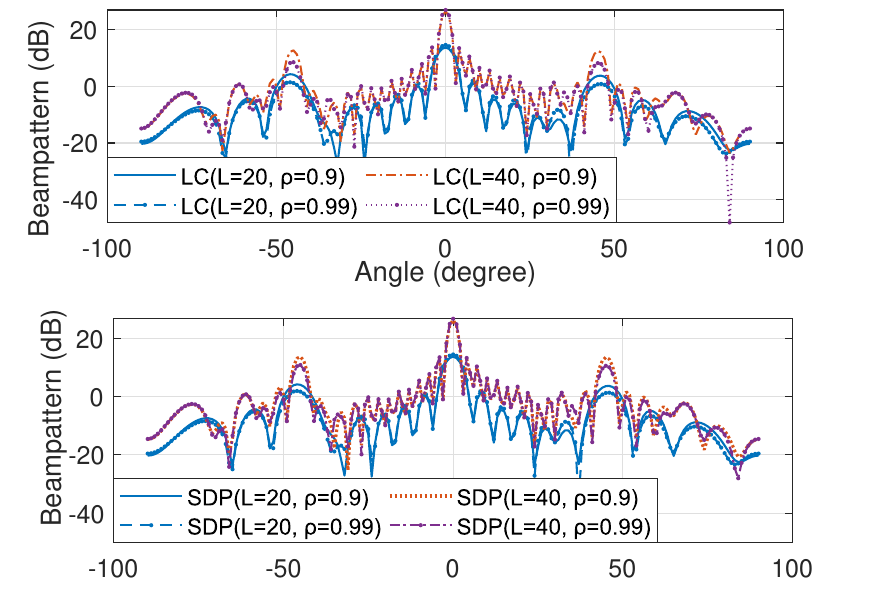}\\
  \caption{ Beampattern with respect to $L$ and $\rho$.}
\end{figure}

Finally in Fig. 5 we show the beampattern designed by the proposed algorithms. Especially both beampattern designed are almost the same and provide a higher beampattern gain at the angle of targets with different reflecting elements. Employing more IRS reflecting elements can not only provide higher beampattern gain to the targets, but also can make the designed beampattern a higher resolution to point the targets more accurate. In Fig. 6, the effect of the trade off factor $\rho$ on the beampattern is shown when $\rho$  are 0.9 and 0.99, the result show that when the $\rho$ increases, meaning that the resource allocation for sensing decreases, the resulted beampattern appears to decrease.

\section{Conclusion}
In this paper, we have explored the application of IRS for ISWPT. An optimization problem has been designed, where a trade-off was introduced to schedule the resources for energy harvesting and sensing, so as to strike a designed performance trade-off between the two functions. Two algorithms, namely the SDP and LC algorithms, have been proposed to efficiently solve the non-convex optimization problem. Our studied and simulation results have demonstrated  that the proposed algorithms converge fast and provide significant performance gains over the legacy methods, while the employment of IRS brings significant benefit to the ISWPT systems.

\balance
\bibliographystyle{ieeetr}

\bibliography{paper_draft}

\begin{thebibliography}{10}

\bibitem{8999605}
F.~Liu, C.~Masouros, A.~P. Petropulu, H.~Griffiths, and L.~Hanzo, ``Joint radar
  and communication design: Applications, state-of-the-art, and the road
  ahead,'' {\em IEEE Trans. Commun.}, vol.~68, no.~6, pp.~3834--3862, Jun.
  2020.

\bibitem{8187650}
Z.~Chu, F.~Zhou, Z.~Zhu, R.~Q. Hu, and P.~Xiao, ``Wireless powered sensor
  networks for internet of things: Maximum throughput and optimal power
  allocation,'' {\em IEEE Internet Things J.}, vol.~5, no.~1, pp.~310--321,
  Feb. 2018.

\bibitem{10077119}
R.~Liu, M.~Li, H.~Luo, Q.~Liu, and A.~L. Swindlehurst, ``Integrated sensing and
  communication with reconfigurable intelligent surfaces: Opportunities,
  applications, and future directions,'' {\em IEEE Wireless Commun.}, vol.~30,
  no.~1, pp.~50--57, Feb. 2023.

\bibitem{9566705}
Z.~Zhu, Z.~Li, Z.~Chu, G.~Sun, W.~Hao, P.~Liu, and I.~Lee, ``Resource
  allocation for intelligent reflecting surface assisted wireless powered {IoT}
  systems with power splitting,'' {\em IEEE Trans. Wireless Commun.}, vol.~21,
  no.~5, pp.~2987--2998, May 2022.

\bibitem{9838546}
Z.~Zhu, Z.~Li, Z.~Chu, G.~Sun, W.~Hao, P.~Xiao, and I.~Lee, ``Resource
  allocation for {IRS} assisted {mmWave} integrated sensing and communication
  systems,'' in {\em Proc. ICC 2022}, pp.~2333--2338, Seoul, Korea, 2022.

\bibitem{10124870}
Z.~Zhu, Z.~Li, Z.~Chu, Q.~Wu, J.~Liang, Y.~Xiao, P.~Liu, and I.~Lee,
  ``Intelligent reflecting surface-assisted wireless powered heterogeneous
  networks,'' {\em Accepted by IEEE Trans. Wireless Commun.}, pp.~1--1, Early
  Access, 2023.

\bibitem{8288677}
F.~Liu, C.~Masouros, A.~Li, H.~Sun, and L.~Hanzo, ``{MU}-{MIMO} communications
  with {MIMO} radar: From co-existence to joint transmission,'' {\em IEEE
  Trans. Wireless Commun.}, vol.~17, no.~4, pp.~2755--2770, Apr. 2018.

\bibitem{9928277}
Q.~Yang, H.~Zhang, and B.~Wang, ``Beamforming design for integrated sensing and
  wireless power transfer systems,'' {\em IEEE Commun. Lett.}, vol.~27, no.~2,
  pp.~600--604, Feb. 2023.

\bibitem{9349191}
H.~Yang, X.~Yuan, J.~Fang, and Y.-C. Liang, ``Reconfigurable intelligent
  surface aided constant-envelope wireless power transfer,'' {\em IEEE Trans.
  Signal Process.}, vol.~69, pp.~1347--1361, Feb. 2021.

\bibitem{9771801}
X.~Song, D.~Zhao, H.~Hua, T.~X. Han, X.~Yang, and J.~Xu, ``Joint transmit and
  reflective beamforming for {IRS}-assisted integrated sensing and
  communication,'' in {\em Proc. 2022 IEEE WCNC}, pp.~189--194, Austin, TX,
  USA, 2022.

\bibitem{9769997}
R.~Liu, M.~Li, Y.~Liu, Q.~Wu, and Q.~Liu, ``Joint transmit waveform and passive
  beamforming design for {RIS}-aided {DFRC} systems,'' {\em IEEE J. Sel. Top.
  Signal Process.}, vol.~16, no.~5, pp.~995--1010, Aug. 2022.

\bibitem{chen2022isac}
Y.~Chen, H.~Hua, and J.~Xu, ``{ISAC} meets {SWIPT}: Multi-functional wireless
  systems integrating sensing, communication, and powering,'' arXiv:
  2211.10605, 2022.

\bibitem{9559408}
Z.~Chu, Z.~Zhu, X.~Li, F.~Zhou, L.~Zhen, and N.~Al-Dhahir, ``Resource
  allocation for {IRS}-assisted wireless-powered {FDMA} {IoT} networks,'' {\em
  IEEE Internet Things J.}, vol.~9, no.~11, pp.~8774--8785, Jun. 2022.

\bibitem{Boyd2004Convex}
S.~Boyd and L.~Vandenberghe, {\em Convex Optimization}.
\newblock Cambridge University Press, 2004.

\end{thebibliography}
\end{document}